\documentclass[twocolumn,english,superscriptaddress]{revtex4}
\usepackage{graphicx}

\usepackage{babel}

\begin{document}

\title{Mini-jet thermalization and diffusion of transverse momentum correlation
in high-energy heavy-ion collisions}

\author{Long-gang Pang}

\affiliation{Interdisciplinary Center for Theoretical Study and Department of
Modern Physics, University of Science and Technology of China, Hefei
230026, China}

\author{Qun Wang}

\affiliation{Interdisciplinary Center for Theoretical Study and Department of
Modern Physics, University of Science and Technology of China, Hefei
230026, China}

\author{Xin-Nian Wang}

\affiliation{Nuclear Science Division, MS 70R0319, Lawrence Berkeley
National Laboratory, Berkeley, CA 94720} 
\affiliation{Institut f\"ur Theoretische Physik, Johann Wolfgang
Goethe-Universit\"at, Max-von-Laue-Str. 1, D-60438 Frankfurt am
Main, Germany}

\author{Rong Xu}

\affiliation{Institute of Particle Physics, Huazhong Normal
University, Wuhan 430079, China} 
\affiliation{Institut f\"ur
Theoretische Physik, Johann Wolfgang Goethe-Universit\"at,
Max-von-Laue-Str. 1, D-60438 Frankfurt am Main, Germany}

\begin{abstract}
Transverse momentum correlation in azimuthal angle of produced
hadrons due to mini-jets are studied first within the HIJING Monte
Carlo model in high-energy heavy-ion collisions. Jet quenching in
the early stage of thermalization is shown to lead to significant
diffusion (broadening) of the correlation. Evolution of the
transverse momentum density fluctuation that gives rise to such
correlation in azimuthal angle in the later
stage of heavy-ion collisions is further investigated within a
linearized diffusion-like equation and is shown to be determined by
the shear viscosity of the evolving dense matter. Such a diffusion
equation for the transverse momentum fluctuation is solved with
initial values given by HIJING and together with the hydrodynamic
equation for the bulk medium. The final transverse momentum
correlation in azimuthal angle is calculated along the freeze-out
hyper-surface and is found further diffused for larger values
of shear viscosity to entropy density ratio $\eta/s \sim 0.2-0.4$.
Therefore the final transverse momentum correlation in azimuthal
angle can be used to study the thermalization of mini-jets in the
early stage of heavy-ion collisions and the viscous effect in the hydrodynamic
evolution of  the strongly coupled quark gluon plasma.
\end{abstract}

\maketitle

\textit{1. Introduction.}  In high-energy heavy-ion collisions at the Relativistic Heavy-ion Collider (RHIC),
enormous amount of transverse energy is produced \cite{Back:2000gw,phenix,Adler:2001yq} via hard and semi-hard
collisions of beam partons from each colliding nuclei \cite{hwa, kajantie,Eskola:1988yh,Wang:1991xy,Eskola:1995bp}. Final state interaction
among these produced partons during the early stage of heavy-ion collisions will lead to a
thermalized (or partially thermalized) quark-gluon plasma (QGP) which will undergo further hydrodynamic
evolution and expansion before hadronization and freeze-out. The strong elliptic flow as measured in
non-central $Au+Au$ collisions at the RHIC energy \cite{star} indicates indeed the early thermalization
of the dense partonic system \cite{heinz} before hydrodynamic expansion takes place and the
momentum anisotropy of the medium partons is developed. The phenomenal jet quenching as observed in the
suppression of large transverse momentum single hadron spectra \cite{phenix2} and back-to-back hadron
correlation \cite{star-dijet} also points to strong interaction between energetic partons and
the bulk medium in the early stage of collisions. Extrapolation of the jet quenching picture to low
and intermediate energy partons leads naturally to a picture  of strong parton interaction in the
bulk medium, a prerequisite for parton thermalization during the early stage of the formation of QGP in heavy-ion collisions.

Jets in high-energy collisions are characterized experimentally by a collimated distribution of hadrons in the direction of
the underlying energetic parton and are measured as such in the calorimetric study of final transverse energy distribution.
Such a calorimetric study of jets becomes problematic and eventually impossible when the transverse energy
of jets becomes increasingly small and comparable to the background of the underlying soft interaction.
However, the characteristic particle correlation within a jet and back-to-back correlation from di-jets should
still be present in the azimuthal angle correlation of produced hadrons if mini-jets dominate the hadron
production mechanism \cite{xnw-jetcor}. Such correlation due to jet and mini-jet production in heavy-ion
collisions has been used to study jet quenching \cite{star-dijet} and possible medium response to jet
propagation \cite{phenix-cone} at RHIC. Extension of such study of correlation to the bulk matter should
shed light on mini-jet thermalization throughout the evolution of the dense matter. Since correlations
in a thermally equilibrated medium  also carry information about the dynamics in an interacting
QGP \cite{Stephanov:1999zu,Jeon:2000wg,Stephanov:2001zj},
study of the transverse momentum correlation in relativistic heavy ion collisions in principle can also provide 
constraints on the shear viscosity of the interacting dense medium \cite{Aziz:2004qu,Gavin:2006xd,Wang:2009zz}.
Furthermore, experimental study of back-to-back hadron correlation in azimuthal angle can also provide 
information on the initial parton production mechanism and whether mono-jet production in the gluon 
saturation model \cite{kharzeev} dominates.

In this paper, we propose and study the transverse momentum
correlation in azimuthal angle as a direct measure of both the early
parton thermalization and later viscous hydrodynamics evolution 
in high-energy heavy-ion collisions. We first
calculate the transverse momentum correlation in azimuthal angle
within the HIJING \cite{hijing} Monte Carlo model and study the effect
of early parton interaction through jet quenching mechanism. We will
further study the evolution of transverse momentum fluctuation and
correlation in azimuthal angle within the framework of viscous
hydrodynamics and the diffusion of the transverse momentum
correlation due to shear viscosity of the bulk matter.

\textit{2. Transverse momentum correlation.} Instead of hadron multiplicity correlation, 
we will focus on the transverse momentum correlation for the purpose of our study, 
which should be less sensitive to the hadronization process and can be directly related 
to the momentum density fluctuation in an diffusion equation
derived from viscous hydrodynamics \cite{Aziz:2004qu,Gavin:2006xd} 
as we will show later. We integrate over a finite size of
psudo-rapidity $|\eta| <1.0$ and discretize the azimuthal
angle $\phi$ into a number of bins whose size $\delta\phi\ll 1$ should be much smaller than the
typical size of a jet. The transverse momentum correlation between two bins centered at
azimuthal angle $\phi_{1}$ and $\phi_{2}$ is defined as
\begin{eqnarray}
C_{T} (\phi_{1},\phi_{2})& = & \frac{\left\langle \delta E_{T1} \delta E_{T2} \right\rangle}
{\left\langle E_{T1}\right\rangle \left\langle E_{T2}\right\rangle}\, ,
\label{eq:correlation}
\end{eqnarray}
where $\delta E_{Tj}=E_{Tj}-\langle E_{Tj}\rangle$ ($j=1,2$) are transverse momentum fluctuations
in the $j$-th bin, $E_{Tj}=\sum_i p_{Ti}\theta(\delta\phi/2-|\phi_{i}-\phi_{j}|)$
are the total transverse momenta in each bin, and the average is taken over events.

\begin{figure}
\caption{\label{fig:c12} The transverse momentum correlation $C_{T}(\phi_1,\phi_2)$
from the HIJING Monte Carlo model in $Au+Au$ central collisions
at $\sqrt{s}=200$ GeV with (solid) and without (dashed)  jet quenching.
In central collisions, $C_{T}(\phi_1,\phi_2)$ only depends on $\varphi=\phi_1-\phi_2$.
The transverse momentum fluctuation in the same bin at $\varphi=0$ is subtracted from the correlation.}
\includegraphics[scale=0.4]{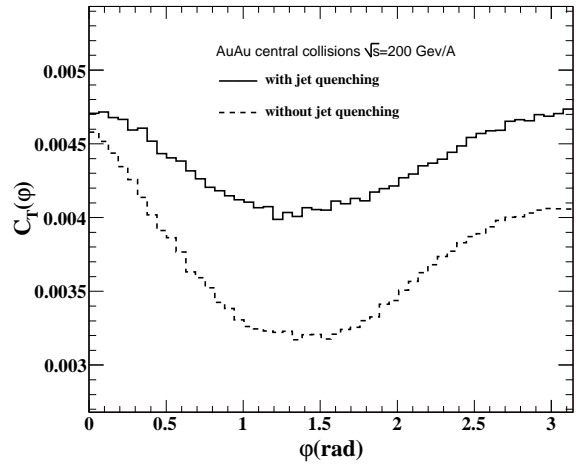}
\end{figure}

We will first study the above transverse momentum correlation using the HIJING Monte Carlo model \cite{hijing},
which is based on a two-component description of soft and semi-hard interactions. Mini-jets contribute to
a significant fraction of hadron production at the RHIC energy. Therefore, transverse momentum
correlation $C_{T}$, as defined in Eq.~(\ref{eq:correlation}),
should reflect that of hadrons from mini-jets as shown
in Fig.~\ref{fig:c12} as the dashed line for central $Au+Au$
collisions at the RHIC energy without jet quenching. The transverse momentum correlation
is similar to two-hadron correlation \cite{xnw-jetcor} with finite  transverse momentum
cut-off. It contains both near-side correlation due to hadron production from  a single mini-jet and
back-side correlation due to hadron production from two back-to-back mini-jets. Final-state
interaction in HIJING is modeled through a schematic parton energy loss mechanism, which will
lead to a broadening or diffusion of the transverse momentum correlation as shown by
the solid line in Fig.~\ref{fig:c12}. Since the parton number density decreases very fast due to longitudinal
expansion, mini-jet interaction is the strongest in the early stage and therefore jet quenching
represents  an early thermalization process for mini-jets in heavy-ion collisions. This  will
lead to significant modification of the characteristic correlation due to mini-jets.
Therefore, the transverse momentum correlation in heavy-ion collisions can be used as
a direct measure of mini-jet thermalization. Since the HIJING results without jet quenching is
equivalent to that of a superposition of independent $p+p$ collisions, one therefore should compare
heavy-ion collisions to $p+p$ to study the degree of mini-jet thermalization. As
illustrated by the HIJING results in Fig. \ref{fig:c12}, early parton interaction leads to a significant
diffusion of the near-side correlation while the shape of the away-side correlation remains
almost intact since it is mainly controlled by the momentum conservation.

In non-central heavy-ion collisions, the transverse hydrodynamic expansion after the initial thermalization leads to
a strong azimuthal asymmetry of hadron distribution or elliptic flow \cite{Ollitrault:1992bk,star} induced by
the asymmetry of the overlapping geometry and therefore will have similar two-hadron correlations as that 
in Fig.~\ref{fig:c12} due to mini-jets.  But such correlation is for the averaged multiplicity and with respect to
the nuclear reaction plane. This is different from the correlation in Eq.~(\ref{eq:correlation}), which focuses
on correlation of fluctuation from event to event. Nevertheless, there might exist contributions from elliptic flow
to the transverse momentum correlation of our interest. Selection of central collisions will therefore
help minimize the influence of the elliptic flow, which can be further reduced by averaging the
transverse momentum correlation $C_{T}(\phi_{1},\phi_{2})$ over all possible pairs of bins with fixed
width $\delta\phi$. In central collisions, $C_T$ only depends on
$\varphi=\phi_1-\phi_2$, so we can use the notation $C_{T}(\varphi)=C_{T}(\phi_1-\phi_2)$.

\textit{3. Diffusion equation.} After the initial (or partial) thermalization, we can
assume that the system will obey the Bjorken scaling and  can be described by the
cylindrical coordinates $X^{\mu}=[\tau,\eta,r,\phi]$ with metrics $g_{\mu\nu}=\mathrm{diag}(-1,\tau^{2},1,r^{2})$,
where $\tau\equiv \sqrt{t^{2}-z^{2}}$ is  the proper time, $\eta$ the spatial rapidity, $r$ the transverse radius and
$\phi$ the azimuthal angle in transverse plane. We will only consider the central rapidity region in central
collisions and therefore the flow velocity can be cast in the simple form $u^{\mu}=\gamma_{T}[1,0,v_{T},0]$,
where $\gamma_{T}=1/\sqrt{1-v_{T}^{2}}$ and $v_{T}=v_{T}(\tau,r,\phi)$ is independent of
the spatial rapidity $\eta$ in the Bjorken scaling scenario. Under this cylindrical system, we can neglect
the diagonal components of the shear stress tensor $\pi^{rr}$, $\pi^{\tau\tau}$ and $\pi^{\phi\phi}$ as compared
to their ideal counterpart $T_{0}^{rr}$, $T_{0}^{\tau\tau}$ and $T_{0}^{\phi\phi}$, and consider
the only off-diagonal component $T^{\phi r}=\pi^{\phi r}=-\eta_{s}\gamma_{T}^{3}\partial_{\phi}v_{T}/r^{2}$
of the shear stress tensor in the $r$-component of the fluid equations, $\nabla_{\mu}T^{\mu r}=0$,
which can be written as
\begin{eqnarray}
\frac{\eta_{s}}{r^{2}}\frac{\partial}{\partial\phi}\left(\gamma_{T}^{3}
\frac{\partial v_{T}}{\partial\phi}\right)
& = & \frac{1}{\tau}\frac{\partial}{\partial\tau}\left[\tau(\epsilon+P)\gamma_{T}^{2}v_{T}\right]
+\frac{\partial P}{\partial r} \nonumber \\
& &\hspace{-0.6in} +\frac{1}{r}(\epsilon+P)\gamma_{T}^{2}v_{T}^{2}
+\frac{\partial}{\partial r}\left[(\epsilon+P)\gamma_{T}^{2}v_{T}^{2}\right],
\label{eq:fluid-eq}
\end{eqnarray}
where $\eta_{s}$ denotes the shear viscosity and should be independent of the azimuthal angle $\phi$ in the
case of central heavy-ion collisions, and $\epsilon$ and $P$ are energy density and pressure, respectively.

As shown by the HIJING results in Fig.~\ref{fig:c12}, the magnitude
of the transverse momentum correlation due to mini-jets is quite
small. Therefore, in the study of the evolution of the transverse
momentum correlation, the fluctuation of energy-momentum density as
a result of the correlation can be considered as a perturbation in a
near-equilibrium system. Such a perturbation can be characterized by
the perturbation  $\delta v_{T}$ in the transverse flow velocity
$v_{T}\rightarrow v_{T}+\delta v_{T}$. Neglecting fluctuation in
$\epsilon$ and $P$ and considering central collisions where $v_{T}$
is independent of $\phi$, the equation for the corresponding
fluctuation in the transverse flow velocity can be derived from Eq.
(\ref{eq:fluid-eq}) as,
\begin{eqnarray}
\frac{\eta_{s}}{r^{2}}\gamma_{T}^{3}\frac{\partial^{2}\delta
v_{T}}{\partial\phi^{2}} & = &
\frac{1}{\tau}\frac{\partial}{\partial\tau}
\left[\tau(\epsilon+P)\gamma_{T}^{4}(1+v_{T}^{2})\delta v_{T}\right]\nonumber\\
&  & +\frac{2}{r}\frac{\partial}{\partial
r}\left[r(\epsilon+P)\gamma_{T}^{4}v_{T}\delta v_{T}\right].
\label{eq:fluct-eq03}
\end{eqnarray}
The above is a diffusion-like equation, which can be seen
by observing that in the co-moving frame where $v_T=0$, the above
equation becomes the standard diffusion equation $\partial
(\tau\delta v_T)/\partial\tau = D \partial ^2(\tau\delta v_T)/\partial\phi ^2$ with the 
diffusion constant $D=\eta_s/(Tsr^2)$ if one neglects 
the time-dependence of the enthalpy density $sT$.

With a given initial distribution of $\delta v_T$ one can solve the above diffusion
equation and then calculate the final transverse energy fluctuation at the freeze-out surface,
\begin{eqnarray}
\delta E_T & = & \int d\Sigma_{\mu}u_\nu\delta T^{\nu\mu}
\nonumber\\
&=&\Delta\eta \int \left(v_{T}-\frac{\partial\tau}{\partial r}\right)\gamma_{T} 
\delta g_T r\tau drd\phi ,
\label{eq:et1}
\end{eqnarray}
where $\Delta \eta$ is the spatial rapidity interval,
\begin{equation}
d\Sigma_{\mu}=(-1,0,\frac{\partial\tau}{\partial r},0)r\tau drd\phi
d\eta
\end{equation}
is the freeze-out hyper-surface area element, and
\begin{equation}
\delta g_T(\tau ,r,\phi )= 
\gamma_{T}^{2}(\epsilon+P)\delta v_{T}
\label{eq:et2}
\end{equation}
is the fluctuation in transverse momentum density.
One can then compute the azimuthal correlation of the
transverse energy at freeze-out,
\begin{eqnarray}
&&C_T(\varphi ) = \frac{(\Delta\eta)^{2}(\delta \phi)^2}{\langle
E_{T}(\phi)\rangle ^2_{\mathrm{out}}}
\int _{\mathrm{out}}r_{1}\tau_{1}dr_{1}\int _{\mathrm{out}}r_{2}\tau_{2}dr_{2}\nonumber \\
&& \times\left\langle \delta g_{T}(\tau_{1}(r_{1}),r_{1},\phi)
\delta g_{T}(\tau_{2}(r_{2}),r_{2},\phi+\varphi)\right\rangle
_{\mathrm{out}}, \label{eq:en-col}
\end{eqnarray}
where the average should be over an ensemble of initial conditions for the spatial
distribution of $\delta v_{T}$ and
\begin{eqnarray}
\langle E_{T}\rangle _{\mathrm{out}} & = & \int
_{\mathrm{out}}d\Sigma_{\mu}u_\nu
T^{\nu\mu} \nonumber \\
&=& \Delta\eta \delta \phi \int _{\mathrm{out}}\left(1-v_{T}\frac{\partial \tau }{\partial r } \right)
\gamma _T \epsilon\; r\tau dr . \label{eq:et}
\end{eqnarray}
is the averaged transverse energy per bin.

To solve the diffusion equation (\ref{eq:fluct-eq03}), one has to
provide the initial condition for $\delta v_T$ as well as the space-time
evolution of $v_T$, $\epsilon$ and $P$ as functions of $\tau$ and
$r$.  To obtain the space-time dependence of $v_T$, $\epsilon$
and $P$, we solve the ideal hydrodynamic equations in (2+1)-dimension numerically
\cite{Kolb:2000sd,Kolb:2003dz} with the initial transverse velocity $v_T(\tau_0)=0$
and initial temperature of
$T=359$ MeV (corresponding to energy density $\epsilon _0=30.0$
GeV/fm$^3$) at the center of the overlap region of central $Au+Au$
collisions at an initial time $\tau_0=0.6$ fm/c. The initial transverse distributions 
of $\epsilon$ and $P$ are given by the Glauber model for nuclear collisions with
the Woods-Saxon nuclear distribution. With such initial conditions for (2+1)-d ideal 
hydrodynamic equations, the final transverse energy $dE_{T}/d\eta = 667$ GeV,  
calculated at a freeze-out temperature $T_{f}=128 $ MeV (or energy density 0.075 GeV/fm$^{3}$)
in central (impact-parameter $b=0$) $Au+Au$ collisions at $\sqrt{s}=200$ GeV/n, 
is consistent with the experimental value
$dE_{T}/d\eta=606\pm 35$ GeV for 0-5\% central $Au+Au$ collisions \cite{Adler:2004zn},
which is also close to the value of HIJING result $dE_{T}/d\eta=750$ GeV.
Note that the transverse energy as defined in Eq.~(1), Eqs.~(\ref{eq:et1})--(\ref{eq:et2}) and (\ref{eq:et}) 
is slightly different from the calorimetric energy measured in experiments.
However, the differences are small if one considers only hadrons in the central rapidity region.

We will use the HIJING model to estimate the initial condition for $\delta v_T$.
We use the event ensemble for $\delta E_T$ generated by HIJING (with jet quenching) 
to extract the angular distribution for $\delta v_{T}(\tau _0)$ at the initial time according to
Eqs.~(\ref{eq:et1}) and (\ref{eq:et2}), assuming $\delta v_{T}(\tau _0)$ is independent 
of $r$ in the overlapped region for simplicity. This procedure of extracting the
velocity fluctuation from HIJING, however, must rely on our knowledge of the
initial values of energy density $\epsilon$ and pressure $P$ at the start of the hydrodynamic
evolution when the system is in complete or partial thermalization at $\tau_{0}$.
The initial values of energy density and pressure that we use for solving the ideal hydrodynamic 
equation, however, would give twice the total transverse energy from HIJING if we assume 
a free-streaming freeze-out at the initial time $\tau_{0}$. It is the PV work by the Bjorken 
longitudinal expansion in the (2+1)-d ideal hydrodynamics that causes the total
transverse energy at the final freeze-out to be reduced by half. To match this initial condition
for the ideal hydrodynamics, we will then multiply $\delta E_T$ from the HIJING events by a factor
of 2 so that the transverse momentum correlation evaluated from Eqs.~(\ref{eq:et1})--(\ref{eq:et})
with a free-streaming freeze-out at the initial time $\tau_{0}$ becomes identical to the
HIJING result (with jet quenching) as shown in Fig.~1.

\begin{figure}
\caption{\label{fig:fm-frz}(color online) The transverse momentum
correlation evolved from the diffusion equation as a function of
azimuthal angle difference $\varphi$ with different values of
$\eta_{s}/s$ on the freeze-out hyper-surface. The EOS is that with
the first order phase transition or EOSQ. }
\includegraphics[scale=0.4]{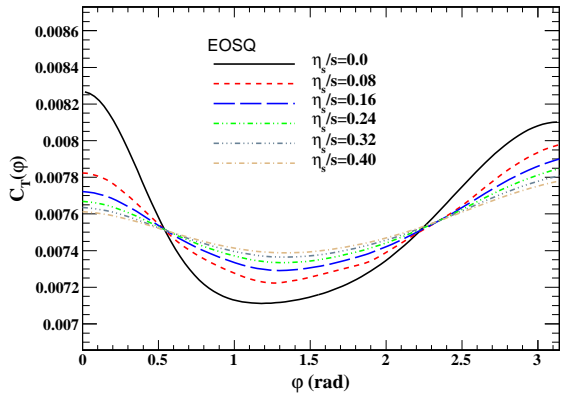}
\end{figure}

In the calculation of the transverse momentum correlation $C_{T}(\varphi)$ from HIJING
events as shown in Fig.~\ref{fig:c12}, we have removed the contribution  from transverse
momentum fluctuation $\sum_{i} \langle E_{Ti}^2 \rangle
-\langle E_{Ti} \rangle ^2$ in the same bin at $\varphi =0$. Such fluctuation should also exist in the
initial condition for $\delta v_{T}$ from HIJING events and will influence the transverse
energy correlation at $\varphi\neq 0$ at a later time due to the diffusion according to Eq.~(\ref{eq:fluct-eq03}).
We will use the following procedure to remove the influence of the same-bin transverse momentum 
fluctuation on the final transverse momentum correlation through the diffusion equation.
We first generate an event ensemble of $\delta v_T$ whose distribution in $\phi$ is random and the initial
transverse momentum correlations all vanish except at the first bin ($\varphi=0$). The
magnitude of the random $\delta v_T$ distribution is adjusted such that the initial transverse correlation
in the first bin at $\varphi=0$ matches that given by the HIJING events. We then solve the diffusion equation
(\ref{eq:fluct-eq03}) with such a random distribution of $\delta v_{T}$ and calculate the event-averaged
transverse momentum correlation $C_{T}^{(0)}(\varphi)$ at the same final freeze-out surface. Such correlation
will be subtracted from the results with the HIJING initial condition to give us the final transverse
energy correlation without the influence of the same-bin fluctuation.

Shown in Fig.~\ref{fig:fm-frz} are the final event-averaged
transverse momentum correlations $C_{T}(\varphi)$ from the solution of
the diffusion equation (\ref{eq:fluct-eq03}) with different values
of the shear viscosity to entropy density ratio $\eta_{s}/s$. The
azimuthal dependence of $\delta g_{T}$ from the solution of
Eq.~(\ref{eq:fluct-eq03}) will be independent of time for zero
viscosity $\eta_{s}=0$. However, its normalization will change because of
the longitudinal and transverse expansion. During the evolution, both the
transverse momentum fluctuation and the averaged transverse energy will
decrease with time. However, the latter decreases faster as governed by the
ideal hydrodynamics. This leads to the increase of the magnitude of the
transverse correlation $C_{T}(\varphi)$ due to hydrodynamic expansion
even for ideal fluid with $\eta_{s}/s=0$ while the shape of the correlation remains the same.
This is consistent with the observation that the magnitude of the transverse
energy correlation from HIJING with jet quenching is larger than that without as
shown in Fig.~1.

As one can see, the transverse momentum correlation becomes diffused
as one increases the value of shear viscosity to entropy density
ratio, which characterizes the momentum exchange among nearby fluid
elements in different bins of azimuthal angle. This is consistent
with the HIJING results with and without jet quenching since jet
quenching happens at the earliest stage of heavy-ion collisions
during the initial thermalization process where the shear viscosity
is the largest.  The diffusion of the correlation in the near-side
due to the viscous evolution is much stronger than the away-side.
We have also studied the dependence of the final
transverse momentum correlation on the equation of state (EOS). 
We find that the shape of the final correlation is
not very sensitive to the EOS, so we present in Fig.
\ref{fig:fm-frz} only the result of the EOS with the first order phase
transition or EOSQ. 


\textit{4. Conclusions and discussions.} We have studied transverse
energy correlation in azimuthal angle and its evolution in
high-energy heavy-ion collisions. We show that such correlation will
have a characteristic structure of two Gaussian peaks in high-energy
heavy-ion collisions at the RHIC energy and above due to the
dominance of mini-jets in the initial parton production. We find
that within the HIJING Monte Carlo model, the transverse energy
correlation will be significantly broadened or diffused by jet
quenching in the early stage of thermalization in heavy-ion
collisions. Assuming that the residual correlation still exists in
the nearly-equilibrated system, we derived a diffusion-like equation
for the local transverse energy density fluctuation by taking
linearized approximation of the viscous fluid equations. With the
initial condition as given by the HIJING results, we solved such a
diffusion equation in combination with the space-time evolution of
the averaged thermodynamic variables and flow velocity as given by
the ideal hydrodynamic equations. The final transverse energy
correlation of produced hadrons averaged over many events is found
to be diffused because of the shear viscosity. For larger values of
the shear viscosity to entropy density ratio, $\eta_{s}/s\sim
0.2-0.4$, the diffusion is as significant as that caused by jet
quenching. One can therefore essentially study the degree of mini-jet
thermalization in heavy-ion collisions as well as the viscous effect
of the strongly coupled QGP \cite{Gyulassy:2004zy,Policastro:2001yc} by
measuring the transverse momentum correlation in heavy-ion collisions
in comparison with that in $p+p$ and $p+A$.

The correlations of transverse momentum per particle  in both pseudo-rapidity and azimuthal angle have
been measured by the STAR experiment \cite{Adams:2005aw} at RHIC through an
autocorrelation technique which inverts the bin-size dependence of the event-wise mean transverse
momentum fluctuation to extract the bin-bin correlation. Such correlation is different from the correlation of
the transverse momentum which is the total transverse momentum within a bin with given size $\delta\phi$.
Nevertheless the transverse momentum correlation that is inferred in the STAR experiments and its
centrality dependence \cite{Adams:2005aw} might point to medium modification of the jet fragmentation
and hadronization process that could also influence the transverse momentum correlation as studied in this paper.
Such effect beyond the normal jet hadronization and hydrodynamic evolution of the medium will be interesting
to investigate through experimental measurement of the transverse momentum correlation as proposed in this
paper.

We thank Y. F. Wu for insightful discussions that got us interested in this problem, and U. Heinz
for providing the code for solving the ideal hydrodynamic equations. We also thank J. Dunlop and L. Ray
for discussions on experimental measurements of transverse momentum correlation. This work is
supported by NSFC of China under Projects No. 10825523, No. 10675109 and No. 10735040
and MOE of China under Project No. IRT0624, and by the Director, Office of Energy Research,
Office of High Energy and Nuclear Physics, Divisions of Nuclear Physics, of the U.S. Department
of Energy under Contract No. DE-AC02-05CH11231. Q.~Wang is supported in part by '100 talents' project of
CAS. During the completion of this work, R.~Xu was supported in part
by the Helmholtz International Center for FAIR within the framework of the LOEWE program
(Landes-Offensive zur Entwicklung Wissenschaftlich-\"okonomischer Exzellenz) launched by the State of Hesse, Germany.
X.-N. Wang thanks the hospitality of the Institut f\"ur Theoretische Physik, Johann Wolfgang Goethe-Universit\"at and support by the
ExtreMe Matter Institute EMMI in the framework of the Helmholtz Alliance Program of the Helmholtz Association (HA216/EMMI)
during the completion of this work.

\end{document}